\documentclass[doublecol,linenumbers]{epl2}

\title{Partial order induced superconductivity in $Fe^{2+}$ iron.}
\shorttitle{Superconductivity in $Fe^{2+}$ iron.}

\author{Naoum Karchev}

\institute{Department of Physics, University of Sofia, 1164 Sofia, Bulgaria}


\newcommand{\be}{\begin{equation}}
\newcommand{\ee}{\end{equation}}
\newcommand{\bea}{\begin{eqnarray}}
\newcommand{\beaa}{\begin{eqnarray*}}
\newcommand{\eea}{\end{eqnarray}}
\newcommand{\eeaa}{\end{eqnarray*}}

\usepackage{epsfig}
\usepackage{color}
\usepackage{amsmath}
\usepackage{amssymb}
\usepackage{bbm}
\usepackage{graphicx}

\abstract {In this letter,  we address the $Fe^{2+}$ state of iron with six 3d electrons. Five of them are localized with ferromagnetic order, while the sixth one is itinerant antiparallelly to the localized ones. We consider spin-fermion model of 3d electrons and show that one can fix the parameters in the theory so that the calculated magnetization matches the experimentally measured one.
With these parameters in mind we show that the sixth 3d electrons have well defined Fermi surfaces, therefore the material is metal.
Further on we consider an iron prepared by means of an applied external magnetic field upon cooling. We assume that the applied magnetic field is along the magnetic order of localized 3d electrons and antiparallel to the magnetic order of the itinerant sixth electron. Therefore the applied field decreases the Zeeman splitting of the spin-up and spin-down sixth electrons. We focus on the quantum partial order (QPO) state which is obtained when the applied field compensates the Zeeman splitting so that the sixth 3d electrons do not contribute to the magnetism of iron and magnetic order is formed by means of localized 3d electrons. We obtain an effective Hamiltonian for iron in (QPO) state and demonstrate that the spin triplet superconducting state with the $T_{1u}$ gap symmetry may be a ground state.
}
\begin{document}

\maketitle

\section{Introduction}

The objective of the study of ferromagnetic metals (iron, cobalt, nickel) is to explain the coexistence of magnetic order and conductivity in these materials.
At the heart of this phenomenon lies the fact that part of the electrons in the system is localized while others are itinerant.
The theory of the magnetism of localized electrons is well described  by means of Heisenberg model,  while the theory of itinerant electron magnetism is still under debate. The origin of spin polarization of itinerant electrons is studied by Stoner \cite{Stoner}. His theory of itinerant electron magnetism, based on spin-polarized band theory, describes  the magnetic ground state in metals but fails to calculate the Curie temperature. This disadvantage was overcome by Moriya \cite{Moriya} who developed  a theory of spin fluctuations in itinerant electron systems.

The first theory of ferromagnetic metals, with account for itinerant and localized electrons, is presented by  Vonsovsky \cite{Vonsovsky}.  The main idea is that s-electrons in the system are responsible for conductivity while d-electrons are responsible for the magnetism (s-d model). in a more rigorous way the magnetic metals are discussed by T. Kasuya \cite{Kasuya}.  He showed, that both ferro- and antiferromagnetism are possible.

An important advancement in the theory of ferromagnetic metals are the  principles formulated by Zener. The third Zener's principle \cite{Zener51} is that the spin of an incomplete d shell is strongly coupled to the spin of the conduction electrons. This coupling tends to align the spins of the
incomplete d shells in a ferromagnetic manner. Guided by the Zener's principle one can formulate the theory of ferromagnetic metals in terms of d electrons only. To match the experimental results for saturation magnetization in units of $\mu_B$ per lattice site $M=2.217$, one has to consider a model with six d electrons. Five of them  are localized and parallelly oriented while the sixth one, in accordance with Pauli principle, is antiparallelly aligned with respect to localized electrons.  If the sixth electron is localized too, the magnetization per lattice site is 2$\mu_B$. To resolve this shortcoming Zener proposed that the average number of the sixth electrons is less than one per atom. This means that ferromagnetic iron is  multivalent ($Fe^{2+}-Fe^{3+})$. The fact that the major portion of the $Fe$ moment is localized comes from neutron scattering experiments \cite{Lowde56}, and the specific-heat measurements\cite{Hofman56}.

In the present paper we consider the spin-fermion model of ferromagnetism and conductivity of $Fe^{2+}$ iron. The five localized electrons are described by spin $s=5/2$ spin operators, while the sixth electrons are fermions. The spin-fermion exchange is antiferromagnetic. To explain the magnetism and conductivity of the iron we invoke the Mott theory of the insulator-metal transition \cite{Mott68}. It states that if the  kinetic energy of the electron is high enough compared with Coulomb repulsion,  doubly occupied states can be realized and respectively empty states. Then the hopping of the electrons realizes the electric transport and the material is metal. At the same time the doubly occupied and empty states are spin singlets, so that their existence effectively decreases the value of the magnetic moment of the system. One can fix the parameters of fermion model so that the total magnetic moment per atom in units of Bohr magneton to match the experimental measurement.

To explore more precisely the impact of the spin fluctuations in ferromagnetic metals one maps the itinerant electron system onto an effective Heisenberg model with classical spins \cite{EHM1,EHM2,EHM3,EHM4}. With exchange parameters in mind the effective Hamiltonian is used to study the spin fluctuations in the system.

In the present paper we integrate out the fermions and obtain an effective Heisenberg hamiltonian of two spins, an analog of Heisenberg model of ferrimagnets\cite{Neel48}. The important difference is that in the present paper the two spin operators are at one and just the same lattice site.
To study magnetic properties of the iron we borrow technique of calculations utilized for ferrimagnetic systems\cite{Diep97,Karchev15}.

 We proceed studying field-cooled (FC) iron. The material is named field-cooled if, during its preparation, an external magnetic field is applied upon cooling \cite{spinel+,spinelFeCr2S4,spinelCv1,spinel08,spinel++,spinel11b,spinel11a,spinel11c,spinel12a,spinel12b,spinel+1,spinel+2,spinel+3,FeV2O4-14}.
 The experimental results show a notable difference of  the magnetic properties of the FC ferrimagnetic spinel and the normal one below N\'{e}el $T_N$ temperature.

 FC systems possess an important state, the partial order state. Magnetic state  is a partial order state if only part of the electrons in the system give contribution to the magnetic order. It is studied in exactly solvable models \cite{Vaks66,Azaria87,Diep04}, by means of Green's function approach \cite{Diep97} or modified spin-wave theory of magnetism \cite{Karchev15} and utilizing the Monte Carlo method \cite{Azaria87}. Experimentally the partial order is observed in $Gd_2Ti_2O_7$ \cite{POExp04}.

 In the present paper we assume that the applied, during the preparation, magnetic field is along the magnetic order of localized 3d electrons,  therefore antiparallelly to the magnetic order of the itinerant sixth electron. The applied field decreases the Zeeman splitting of spin-up and spin-down sixth electrons . We consider the quantum partial order state when the Zeeman splitting is compensated and the sixth electrons do not contribute the magnetization of iron. The Hamiltonian of the spin-fermion interaction of the sixth electrons, without Zeeman splitting, and transversal fluctuations (magnons) of the localized spins is obtained. We show that this interaction leads to the spin-1  antiparallel p-type superconductivity, and present the dependence of the gap on the parameters in the theory.

 Because the preparation of the field cooled ferrimagnets is well established practice, we hope that there will be no problem in preparation of FC iron and verification of the superconductivity of the iron in the quantum partial order state.

 \section{Ferromagnetism and conductivity of $Fe^{2+}$ iron}

 The Hamiltonian of the 3d electrons of iron in $Fe^{2+}$ state is
 \bea \label{iron11}
 & h & = h_1\,+\,h_2\,+\,h_3\,+\,h_4 \\
 & h_1 &  =-t\sum\limits_{< ij>} \left( c_{i\sigma }^ + c_{j\sigma } + h.c. \right)
  +U\sum\limits_{i} n_{i\uparrow}n_{i\downarrow}-\mu \sum\limits_{i} {n_i} \nonumber \\
& h_2 & = -  J_0^{it}\sum\limits_{<  ij >} {\bf S}_i^{it}\cdot {\bf S}_j^{it} \nonumber \\
& h_3 & =  -  J^l\sum\limits_{  <  ij>} {\bf S}_i^l \cdot {\bf S}_j^l \nonumber \\
& h_4 & = J\sum\limits_{ i  } {\bf S}_i^{it}\cdot {\bf S}_i^l . \nonumber
\eea
The spin operators ${\bf S}_i^l$ are the spin-$5/2$  operators of the localized electrons, $c_{i\sigma }^+$ and $c_{i\sigma }$ ($\sigma=\uparrow,\downarrow$) are creation and annihilation operators for spin-1/2 Fermi operators of itinerant electrons, ${\bf S}_i^{it}$ are the spin operators of the itinerant
electrons with components
\be\label{iron11a}
S^{it}_{ \nu i}=\frac 12\sum\limits_{\sigma\sigma'}c^+_{i\sigma}\tau^{\nu}_{\sigma\sigma'}c^{\phantom +}_{i\sigma'},\ee
where $(\tau^x,\tau^y,\tau^z)$ are Pauli matrices, $n_{i\sigma}=c^+_{i\sigma}c_{i\sigma}$, $n_i=n_{i\uparrow}+n_{\downarrow}$,
$t>0$ is the hopping parameter, $U>0$ is the Coulomb repulsion and $\mu$ is the chemical potential. The parameters $J_0^{it}$, $J^l$ and $J$ are positive and describe the ferromagnetic exchange between itinerant electrons ($J_0^{it}$), between localized electrons ($J^l$) and the antiferromagnetic spin fermion exchange ($J$) . The parameter $J$ is characteristic of intra-atom interaction, while $J^l$ features the exchange between two sites (two atoms). This is why $J$ is much larger then $J^l$. Sums are over all sites of a body centered cubic (bcc) lattice.

To study the effects of hopping $t$ and Coulomb repulsion $U$ we represent the fermi operators $c_{i\sigma }^+,c_{i\sigma }$, spin operators ${\bf S}_i^{it}$ and density operators $n_{i\sigma}=c^+_{i\sigma}c_{i\sigma}$ of itinerant electrons by means of  the Schwinger-bosons ($\varphi_{i,\sigma}, \varphi_{i,\sigma}^+$) and slave fermions ($h_i, h_i^+,d_i,d_i^+$) \cite{Schmeltzer91}, where the fermions have no spin but have charge 1 and -1 respectively, while bosons are charge-less with spin 1/2. In this representation  Coulomb term is quadratic and one can study its impact exactly \cite{Karchev13}.
The spin operators of localized electrons ${\bf S_i^l}(a^+_j,a_j)$ are represented by means of  Holstein-Primakoff representation where $a^+_j,\,a_j$
are Bose fields \cite{supplementary}.

Accounting for the quadratic terms of slave fermions ($h_i, h_i^+,d_i,d_i^+$) in Hamiltonian $h_1$ and $h_4$ we get the expression for the free Hamiltonian of the fermions
\be\label{iron12}
h_0 =   \sum\limits_{k} \left (\varepsilon^d_k d_k^+ d_k + \varepsilon^h_k h_k^+ h_k \right)\ee
At half filling $\mu=U/2$ and dispersions adopt the form
\bea\label{iron13a}
\varepsilon^d_k & = & -t\varepsilon_k +\frac {U}{2}+\frac {s J}{2}\nonumber \\
\varepsilon^h_k & = & t\varepsilon_k+\frac {U}{2}+\frac {s J}{2} \\
\varepsilon_k &  = & -8 \cos \frac {k_x}{2}\cos \frac {k_y}{2}\cos \frac {k_z}{2},\nonumber \eea
where $k$ runs over the first Brillouin zone of a bcc lattice.

An important characteristic of the sixth electrons is the zero-temperature spontaneous magnetization of the electron
\begin{equation}
m=\frac 12 \left(1-<h^+_i h_i>-<d^+_id_i>\right). \label{iron13}
\end{equation}
When in the ground state the lattice site is doubly occupied $(<d^+_id_i>=1)$ or empty $(<h^+_i h_i>=1)$  $m=0$, these states are singlet. When $m$ is maximal ($m=1/2$), therefore $<h^+_i h_i>=<d^+_id_i>=0$, all sites are occupied by one sixth electron.

To feature the magnetic properties of the sixth electrons we introduce the vector ${\bf M}_i$, ${\bf M}_i^2=m^2$. The spin vector of the itinerant electrons (\ref{iron11a}) can be represented in the form
\be\label{iron14} {\bf S}_i^{it}=\frac {1}{2m}{\bf M}_{i}\left(1-h^+_i\,h_i\,-\,
d^+_i\,d_i\right),\ee

We integrate out the slave fermions $d$ and $h$ to obtain the two-spin Hamiltonian of the effective theory of magnetism of iron
 \bea \label{iron15}
 h^{eff} & = & -  J^{it}\sum\limits_{<  ij >} {\bf M}_i\cdot {\bf M}_j
  -  J^l\sum\limits_{  <  ij>} {\bf S}_i^l \cdot {\bf S}_j^l \nonumber \\
& + & J\sum\limits_{ i  } {\bf M}_i\cdot {\bf S}_i^l ,
\eea
where ${\bf M}_i=<{\bf S}_i^{it}>_{d,h}$ and the exchange constant $J^{it}$ is a positive sum of the exchange constant $J_0^{it}$ (\ref{iron11}) and exchange due to interaction of slave fermions and Schwinger-bosons. To proceed we use the technique of calculation implemented in the theory of ferrimagnetism: i) we represent $S_i^l$ and $M_i$ by means of spin $s=5/2$ and spin $m<1/2$ Holstein-Primakoff formulae, ii) we keep only the quadratic terms in the Hamiltonian, and use the Bogoliubov transformation to diagonalize it. The final result for the effective Hamiltonian in terms of Bogoliubov bosons is \cite{supplementary}
\be\label{iron16}
h^{eff} =   \sum\limits_{k} \left (E^{\alpha}_k \alpha_k^+ \alpha_k + E^{\beta}_k \beta_k^+ \beta_k + E^0_k \right),\ee
where $E^{\alpha}_k>0$ for all values of the wave vector ($\bf k$), $E^{\beta}_k$ is a dispersion of the ferromagnetic magnon of the system and
$E^0_k$ is the ground state energy. The spontaneous magnetization of the system $M=M^l+M^{it}$  has  representation
\be\label{iron16a}
M = s-m -\frac 1N \sum\limits_{k}\left(<\alpha_k^+\alpha_k> - <\beta_k^+\beta_k> \right).\ee
At zero temperature $<\alpha_k^+\alpha_k>=0, <\beta_k^+\beta_k>=0$ and the saturated magnetization is $M=s-m$. The experimental result for $M$ is $M=2.217$ in units of Born magneton. For $s=5/2$,the magnetic moment of the itinerant electron is  $m=0.283<0.5$. If $\varepsilon^h_k>0$ and $\varepsilon^d_k>0$, $<h^+_ih_i>=<d^+_id_i>=0$ and $m=0.5$ as follows from equation (\ref{iron14}). The magnetization of the sixth electron is $m<0.5$ if charge carriers fermions $h$ and $d$ have Fermi surfaces ($<h^+_ih_i>+<d^+_id_i>\neq0$). Therefore the material is metal. The theoretical calculations show that  $m=0.283$ if the parameter $(Js+U)/(2t)=1.78$ \cite{supplementary}.

The model (\ref{iron11}) with five localized and one itinerant 3d electrons describes the magnetic and transport properties of $Fe^{2+}$ iron.

\section{Quantum Partial Order and Superconductivity of Field-Cooled $Fe^{2+}$ Iron}

We continue  examining  a field-cooled iron. If, during the preparation, we apply magnetic field  along the magnetic order of localized 3d electrons   the magnetization of these electrons arrive at their saturation, while the Zeeman splitting of itinerants electron decreases. The nontrivial point is that this is true and after  switching off the magnetic field when the process is over \cite{spinelFeCr2S4,spinel+,spinelCv1,spinel08,spinel++,spinel11b,spinel11a,spinel11c}. To account for the effect we include "frozen" magnetic field adding a term $-H \sum\limits_{i} {S^{it \,z}_{i}}$ in the Hamiltonian (\ref{iron11}). Then the dispersions of the slave fermions ($d,h$) adopted the form (\ref{iron13a}):
\bea\label{iron18}
\varepsilon^d_k & = & -t\varepsilon_k +\frac {U}{2}+\frac {s J}{2}-\frac H2 \nonumber \\
\varepsilon^h_k & = & t\varepsilon_k+\frac {U}{2}+\frac {s J}{2}-\frac H2, \eea
where $s=5/2$. We consider a state prepared with applied magnetic field $H=H_{c}=U+sJ$. The density of slave fermions $n^d=n^h=1/2$, therefore the magnetic moment of itinerant electron is $m=0$ (\ref{iron13}). By means of this field one prepares a quantum partial order (QPO) state, when the itinerant sixth 3d electron does not contribute the magnetization of the system and magnetic order is formed by the five localized  3d electrons. In QPO state the itinerant electrons do not form transversal spin fluctuations, and we can represent them by means of  creation and annihilation operators in Hamiltonian (1). The Hamiltonian of iron in QPO state is
\bea\label{iron19}
 h^{QPO} & = & -t\sum\limits_{< ij>} {\left( {c_{i\sigma }^ + c_{j\sigma } + h.c.} \right)}
   -  J^l\sum\limits_{<ij>} {{\bf S_i^l}
\cdot {\bf S_j^l}}\nonumber \\
& + & \sqrt{\frac{s}{2}}J\sum\limits_{i}\left(c_{i\downarrow }^ + c_{i\uparrow }a_i+c_{i\uparrow}^ + c_{i\downarrow }a_i^+\right),
\eea
where $a_i,a_i^+$ are Bose operators from the Holstein-Primakoff representation of spin operators of localized electrons $\bf S_i^l$. In the spin-fermion interaction they are accounted for in linear approximation.

To study the superconductivity of iron, induced by spin fluctuations, we integrate out the bosons and obtain an effective four-fermion interaction.
Then we proceed in the standard way to find the gap function equation.

The sums in Eq.(\ref{iron11}) are over all sites of a body centered cubic (bcc) lattice. The first Brillouin zone of a bcc lattice is quite complicated and it is difficult to integrate over wave vectors. To circumvent this problem we introduce two equivalent simple cubic sub-lattices.
Following the standard procedure one obtains the effective Hamiltonian in the Hartree-Fock approximation
\bea\label{iron22s}
h^{f^4}_{HF} & = & \sum\limits_{k}\left[\Delta^A_k c^{A+}_{\downarrow -k}c^{A+}_{\uparrow k} + \Delta^{A+}_k c^A_{\uparrow k} c^A_{\downarrow -k}\right. \nonumber \\
& + &
\left. \Delta^B_k c^{B+}_{\downarrow -k}c^{B+}_{\uparrow k} + \Delta^{B+}_k c^B_{\uparrow k} c^B_{\downarrow -k}\right], \eea
where the gap functions of the wave vector $k$ are defined by the equations
\bea\label{iron23s}
\Delta^A_k & = & \frac {1}{N}\sum\limits_{p}<c^A_{\uparrow -p}c^A_{\downarrow p}> V_{p-k} \nonumber \\
\\
\Delta^B_k & = & \frac {1}{N}\sum\limits_{p}<c^B_{\uparrow -p}c^B_{\downarrow p}> V_{p-k}. \nonumber \eea
The potential $V(p-k)=V(q)$
\be\label{iron21}
V({\bf q})=\frac {J}{16J^l}\frac {1}{1-\cos^2(\frac {q_x}{2})\cos^2(\frac {q_y}{2})\cos^2(\frac {q_z}{2})}\ee
is the fermion binding potential result of transversal spin fluctuations of the 3d localized electrons. The wave vectors $k$ and $p$  run over the first Brillouin zone of a simple cubic lattice.

The two sublattices are equivalent,  therefore the Hamiltonian should be invariant under the replacement $A \leftrightarrows B$. This is true if
\be\label{iron24s}
\Delta^A = \Delta^B = \Delta .
\ee
We set $\Delta $ in equation (\ref{iron22s}) and by means of the Bogoliubov transformation we rewrite the Hamiltonian in a diagonal form
\be\label{iron25s}
h^{f^4}_{HF} = \sum\limits_{k}E_k \left[f^{A+}_k f^A_k + f^{B+}_k f^B_k -\rho^{A+}_k\rho^A_k -\rho^{B+}_k\rho^B_k\right], \ee
where
\be\label{iron26s}
E_k= \sqrt{\varepsilon^2_k + \Delta^2_k} \ee
Bearing in mind Bogoliubov  transformation we calculate
\bea\label{iron27s}
<c_{\uparrow -k}^A c_{\downarrow k^A}> & = &  <c_{\uparrow -k}^B c_{\downarrow k}^B> \\
& = &
\frac 12 \frac {\Delta_k}{\sqrt{\varepsilon^2_k + \Delta_k^2}}\tanh{\frac {E_k}{2T}} \nonumber \eea
where $T$ is the temperature. At zero temperature the gap equation adopts the form
\be\label{iron28s}
\Delta_{k } = \frac12 \frac 1N \sum\limits_{p} \frac {\Delta_p}{\sqrt{\varepsilon^2_p + \Delta^2_p}}V_(p-k) \ee

The classification for spin-triplet functions $\Delta_k=-\Delta_{-k}$ in the case of bcc lattice \cite{RKS10} inspires to look for a gap in the form with $T_{1u}$ configuration
\be\label{iron20a}
\Delta_k=\Delta\left(\sin k_x+\sin k_y+\sin k_z \right). \ee
We multiply both sides of the equation (\ref{iron28s}) by $\left(\sin k_x+\sin k_y+\sin k_z \right)$ and integrate over the wave vector k. Using  equality
\be\label{iron30}
\int\frac {d^3k}{(2\pi)^3}\left(\sin k_x+\sin k_y+\sin k_z \right)^2 = \frac 32 \ee
we obtain the equation for nonzero gap parameter $\Delta$
\bea\label{iron20}
& & 3 =  \int\frac {d^3k}{(2\pi)^3}\frac {d^3p}{(2\pi)^3}V({\bf p} -{\bf k})\nonumber \\
\\
& & \times\frac {(\sin p_x +\sin p_y +\sin p_z)(\sin k_x +\sin k_y +\sin k_z)}{\sqrt{\frac {t^2}{J^2}\varepsilon_p^2+\frac {\Delta^2}{J^2}(\sin p_x +\sin p_y +\sin p_z)^2 }},\nonumber
\eea
We have used that in QPO state, Hamiltonian (\ref{iron19}), the dispersion of fermions is $t\varepsilon_p$ (\ref{iron13}).

\begin{figure}[!ht]
\epsfxsize=\linewidth
\epsfbox{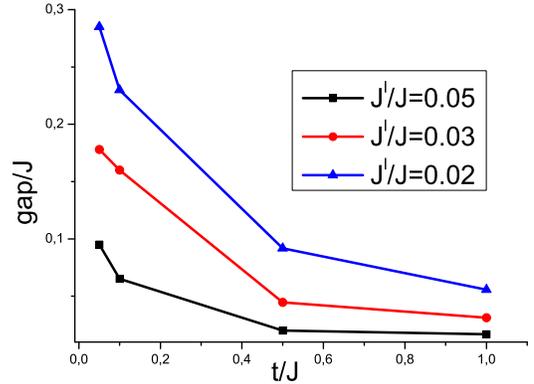} \caption{\,\,The dimensionless $gap/J$ ($\Delta/J$) as a function of dimensionless hoping parameter $t/J$ for different values of the dimensionless parameter $J^l/J$, where $J^l$ is the exchange between the localized d-electrons and $J$ is the exchange between the localized and itinerant d-electrons. The figure shows that superconductivity is suppressed when $t/J$ and $J^l/J$ increase.}\label{Irongap-FeSc}
\end{figure}

The dimensionless gap/J ($\Delta/J$) as a function of dimensionless hoping parameter ($t/J$) is depicted in figure (\ref{Irongap-FeSc}) for different values of the parameter ($J^l/J$). The figure shows that superconductivity is suppressed when $t/J$ and $J^l/J$ increase. It is important to repeat that $J$ is an intra-atomic exchange while $t$ and $J^l$ are exchanges between atoms, hence $J$ is much larger than $t$ and $J^l$. Therefore the small values of $t/J$ and $J^l/J$ are physical relevant. The figures show that this is the case of the factual superconductivity.

\section{Conclusion}

The present paper highlights the possibility of featuring new iron based superconductors.
It is demonstrated that the spin triplet superconducting state with the $T_{1u}$ gap symmetry may be a ground state.

Below Curie temperature ($T_C$) Cooper pairing of fermions is induced by potential (\ref{iron21}) with long range behavior
\be\label{iron22}
V({\bf q}) \approx \frac {1}{{\bf q}^2},\ee
when  ${\bf q} \rightarrow 0 $. Above Curie temperature the spin fluctuations open a gap and one has to replace the potential (\ref{iron21}) in gap equation (\ref{iron20}) by potential with low-momenta aproximation
\be\label{iron23}
V({\bf q}) \approx \frac {1}{{\bf q}^2+\delta} \rightarrow \frac {1}{\delta}.\ee
The gap $\delta$ rapidly increases when temperature increases, which in turn suppresses superconductivity. This is why we think that there is no superconductivity above Curie temperature. A precise analyze of finite temperature properties of the superconductivity is possible after study of the finite temperature properties of ferromagnetic iron in a quantum partial order state.

The same mechanism can be applied in preparation of cobalt based superconductor. The cobalt in $Co^{2+}$ state has seven 3d electrons. One can consider five of them to be localized with ferromagnetic order and the other two itinerant 3d electrons are antiparallel to the localized ones.  We expect two -band superconductivity if the Zeeman splitting energies of the itinerant 3d electrons are equal or two sequencing superconducting states if they are not equal.

The ferromagnetism is incompatible with the s-wave superconductivity. The spin-fermion interaction in these materials leads to Zeeman splitting of itinerant electrons. This is why the superconductivity in these materials is unconventional, near the ferromagnetic quantum critical point, with very low critical temperature. The most famous example of ferromagnetic superconductor is $UGe_2$ \cite{UGe2} with superconductivity confined to ferromagnetic phase, while the most recent materials are the $Eu$-based iron pnictides  $Eu(Fe_0.75Ru0_25)_2As_2$, where the superconductivity coexists with the ferromagnetic order of the $Eu^{2+}$ spins  \cite{EuFe1,EuFe2}.

The superconductivity in the present paper is a result of compensation of Zeeman splitting of itinerant 3d electrons in iron prepared by means of an applied external magnetic field upon a cooling. This reminds us of Jaccarino-Peter (JP) compensation mechanism \cite{JP62}.
For some magnetic metals, the exchange spin-fermion interactions have a negative sign. This  allows for the conduction electron polarization to be canceled by an external magnetic field so that if, in addition, these metals possess phonon-induced attractive electron-electron interaction, superconductivity  occurs in the compensation region. A superconducting state induced by an external magnetic field has been observed in the pseudoternary $Eu-Sn$ molybdenum chalcogenides and explained in terms of the Jaccarino-Peter compensation effect \cite{JP1}.
$CePb_3$ is the first reported heavy-fermion magnetic-field-induced superconductor \cite{JP2}. Such an unusual behavior was interpreted as a manifestation of the Jaccarino-Peter effect.

There is evidence for the Jaccarino-Peter mechanism in the field-induced organic superconductors $\lambda-(BETS)_2FeCl_4$ and $\kappa-(BETS)_2Br_4$ \cite{JP3,JP4,JP5}.

The important point is that in the JP theory the superconductivity appears when the magnetic field is applied and disappears when the magnetic field is switched off.

In the present theory, the magnetic field is switched off when the preparation of field-cooled iron is over. We showed that prepared FC iron in QPO state possesses superconductivity.

The superconductivity of iron is a consequence of nontrivial separation of 3d electrons in $Fe^{2+}$ state. It is due to Pauli principle and permits to prepare a quantum partial order state which in turn is the fundament of superconductivity in FC iron.

The field-cooled $Fe^{2+}$ iron, in a quantum partial order state, is a candidate as a new member of the huge family of iron based superconductors \cite{FeSc1,FeSc2,FeSc3,FeSc4,FeSc5,FeSc6,FeSc7,FeSc8,FeSc9}. There is a theoretical point of view that $FeS$ superconductors are at short distance from Mott transition \cite{FeScTh1,FeScTh2}.


\begin{thebibliography}{0}

\bibitem[*]{byline} Electronic address: naoum@phys.uni-sofia.bg
\bibitem{Stoner}
\Name {E. C. Stoner}
\REVIEW {Proc. Roy Soc. Lonon} {Ser. A 165} {1938} {72}.
\bibitem{Moriya}
\Name {T. Moriya}
\REVIEW {\emph{Spin Fluctuations in Itinerant Electron Magnetism}} {Springer-Verlag, Berlin} {1985}.
\bibitem{Vonsovsky}
\Name {S. V. Vonsovsky}
\REVIEW {\emph{Magnetism} Vol.2} {John Wiley, New York} {1974}.
\bibitem{Kasuya}
\Name {T. Kasuya}
\REVIEW {Prog. Theor. Phys.} {16} {1956} {45}.
\bibitem{Zener51}
\Name {C. Zener}
\REVIEW {Phys. Rev.} {81} {1951} {440}.
\bibitem{Lowde56}
\Name {R. D. Lowde}
\REVIEW {Proc. Roy. Soc. (London)} {235} {1956} {305}.
\bibitem{Hofman56}
\Name {J. A. Hofman, H. Paskin, K. J. Taner, \and R. J. Weis}
\REVIEW {J. Phys. Chem. Solids} {1} {1956} {45}.
\bibitem{Mott68}
\Name {N. F. Mott}
\REVIEW {Rev. Mod. Phys.} {40} {1968} {677}.
\bibitem{EHM1}
\Name {A.I. Liechtenstein, M.I. Katsnelson, and V.A. Gubanov}
\REVIEW {J. Phys. F: Met. Phys.} {14}{1984} {L125}.
\bibitem{EHM2}
\Name {V.P. Antropov, M.I. Katsnelson, B.N. Harmon, M. van Schilfgaarde, \and D. Kusnezov}
\REVIEW {Phys. Rev.} {B 54}{1996} {1019}.
\bibitem{EHM3}
\Name {S.V. Halilov, H. Eschrig, A.Y. Perlov, and P.M. Oppeneer}
\REVIEW {Phys. Rev.} {B 58}{1998} {293}.
\bibitem{EHM4}
\Name {A. Sakuma}
\REVIEW {J. Phys. Soc. Jpn.} {68} {1999} {620}.
\bibitem{Neel48}
\Name {L. N\'{e}el}
\REVIEW {Ann. Phys.} {3} {1948} {137}.
\bibitem{Diep97}
\Name {R. Quartu, H. T. Diep}
\REVIEW {Phys. Rev.} {B 55} {1997} {2975}.
\bibitem{Karchev15}
\Name {N. Karchev}
\REVIEW {J. Mag.Mag Materials} {396} {2015} {77}.
\bibitem{spinelFeCr2S4}
\Name {Zhaorong Yang, Shun Tan, Zhiwen Chen, \and Yuheng Zhang}
\REVIEW {Phys. Rev.} {B 62} {2000} {13872}.
\bibitem{spinel+}
\Name { K. Adachi, T. Suzuki, K. Kato, K. Osaka, M. Takata \and T. Katsufuji}
\REVIEW {Phys. Rev.Lett} {95} {2005} {197202}.
\bibitem {spinelCv1}
\Name {H. D. Zhou, J. Lu, \and C. R. Wiebe}
\REVIEW {Phys. Rev.} {B 76} {2007} {174403}.
\bibitem{spinel08}
\Name {V. O. Garlea, R. Jin, D. Mandrus, B. Roessli, Q. Huang, M. Miller, A. J. Schultz, \and S. E. Nagler}
\REVIEW {Phys. Rev. Lett.} {100} {2008} {066404}.
\bibitem{spinel++}
\Name {S-H. Baek, K-Y. Choi, A. P. Reyes, P. L. Kuhns, N. J. Curro, V. Ramanchandran, N. S. Dalal, H. D. Zhou, \and C. R. Wiebe}
\REVIEW {J. Phys.: Condens. Matter} {20} {2008} {135218}.
\bibitem{spinel11b}
\Name {Kim Myung-Whun, J. S. Kim, T. Katsufuji, and R. K. Kremer}
\REVIEW {Phys. Rev.} {B 83} {2011} { 024403}.
\bibitem{spinel11a}
\Name {A. Kiswandhi, J. S. Brooks, J. Lu, J. Whalen, T. Siegrist, and H. D. Zhou}
\REVIEW {Phys. Rev.} {B 84} {2011} {205138}.
\bibitem{spinel11c}
\Name {A. Kismarahardja, J. S. Brooks, A. Kiswandhi, K. Matsubayashi, R. Yamanaka, Y. Uwatoko, J. Whalen,
T. Siegrist, \and H. D. Zhou}
\REVIEW {Phys. Rev.Lett.} {106} {2011} {056602}.
\bibitem{spinel12a}
\Name {Q. Zhang, K. Singh, F. Guillou, C. Simon, Y. Breard, V. Caignaert, and V. Hardy}
\REVIEW {Phys. Rev.} {B 85} {2012} {054405}.
\bibitem{spinel12b}
\Name { Y. Nii, H. Sagayama, T. Arima, S. Aoyagi, R. Sakai, S. Maki, E. Nishibori, H. Sawa, K. Sugimoto,
H. Ohsumi, \and M. Takata}
\REVIEW {Phys. Rev.} {B 86} {2012} {125142}.
\bibitem{spinel+1}
\Name {Z. H. Huang, X. Luo, S. Lin, Y. N. Huang, L. Hu, L. Zhang, Y. P.Sun}
\REVIEW {Solid State Commun.} {159} {2013} {88}.
\bibitem{spinel+2}
\Name {Z. H. Huang, X. Luo, L. Hu, S. G. Tan, Y. Liu, B. Yuan, J. Chen, W. H. Song, \and Y. P. Sun}
\REVIEW { Journal of Applied Physics} {115} {2014} {034903}.
\bibitem{FeV2O4-14}
\Name {G. J. MacDougall, I. Brodsky, A. A. Aczel, V. O. Garlea, G. E. Granroth, A. D. Christianson, T. Hong,
H. D. Zhou \and S. E. Nagler}
\REVIEW {Phys. Rev.} {B 89} {2014} {224404}.
\bibitem{spinel+3}
\Name {Dina Tobia, Juli$\acute{a}$n Milano, Maria Teresa Causa \and Elin L. Winkler}
\REVIEW {J. Phys.: Condens. Matter}{27} {2015} {016003}.
\bibitem{Vaks66}
\Name {V. G. Vaks, A. I. Larkin, \and Y. N. Ovchinnikov}
\REVIEW {JETP Lett.} {22} {1966} {820}.
\bibitem{Azaria87}
\Name {P. Azaria, H. T. Diep, and H. Giacomini}
\REVIEW {Phys. Rev. Lett.} {59} {1987} {1629}.
\bibitem{Diep04}
\Name {H. T. Diep (ed.)}
\REVIEW {\emph {Frustrated Spin Systems}} {World Scientific, Singapore} {2004}.
\bibitem{POExp04}
\Name {J. R. Stewart, G. Ehlers, A. S. Wills, S. T. Bramwell, \and J. S. Gardner}
\REVIEW {J. Phys.: Condens. Matter} {16} {2004} {L321}.
\bibitem{Schmeltzer91}
\Name {D. Schmeltzer}
\REVIEW {Phys. Rev.} {B 43} {1991} {8650}.
\bibitem{Karchev13}
\Name {N. Karchev}
\REVIEW {Ann. Phys.} {333} {2013} {206}.
\bibitem{supplementary}
\Name {Collection of equations is given in supplementary material with the aim to make more transparent the reading of the manuscript.}
\bibitem{RKS10}
\Name {S. Raghu, S. A. Kivelson, \and D. J. Scalapino}
\REVIEW {Phys. Rev.} {B 81} {2010} {224505}.
\bibitem{UGe2}
\Name{S. Saxena, P. Agarwal, K. Ahilan, F.M. Grosche, R. Haselwimmer,
M. Steiner, E. Pugh, I. Walker, S. Julian, P. Monthoux, G.
Lonzarich, A. Huxley, I. Sheikin, D. Braithwaite, and J. Flouquet}
\REVIEW {Nature (London)} {406} {2001} {58}.
\bibitem{EuFe1}
\Name {Z. Ren, Q. Tao, S. Jiang, C. Feng, C. Wang, J. Dai, G. Cao, and
Z. Xu}
\REVIEW {Phys. Rev. Lett} {102} {2009} {137002}.
\bibitem{EuFe2}
\Name {S. Jiang, H. Xing, G. Xuan, Z. Ren, C. Wang, Z.-A. Xu,
and G. Cao}
\REVIEW {Phys. Rev.} {B 80} {2009} {184514}.
\bibitem{JP62}
\Name {V. Jaccarino and M. Peter}
\REVIEW {Phys. Rev. Lett.} {9} {1962} {290}.
\bibitem{JP1}
\Name {H. W. Meul, C. Rossel, M. Decroux, Ø Fischer, G. Remenyi, and A. Briggs}
\REVIEW {Phys.Rev.Letters} {53} {1984} {497}.
\bibitem{JP2}
\Name {C. L. Lin, J. Teter, J. E. Crow, T. Mihalisin, J. Brooks, A. I. Abou-Aly, and G. R. Stewart}
\REVIEW {Phys. Rev. Lett.} {54} {1985} {2541}.
\bibitem{JP3}
\Name {S. Uji, H. Shinagawa, T. Terashima, T. Yakabe,
Y. Terai, M. Tokumoto, A. Kobayashi, H. Tanaka, and
H. Kobayashi}
\REVIEW {Nature (London)} {410} {2001} {908}.
\bibitem{JP4}
\Name {L. Balicas, J.S. Brooks, K. Storr, S. Uji, M. Tokumoto,
H. Tanaka, H. Kobayashi, A. Kobayashi, V. Barzykin,
and L.P. Gor'kov}
\REVIEW {Phys. Rev. Lett.} {87} {2001} {067002}.
\bibitem{JP5}
\Name {T. Konoike, S. Uji, T. Terashima, M. Nishimura, S. Ya-
suzuka, K. Enomoto, H. Fujiwara, B. Zhang, and
H. Kobayashi}
\REVIEW {Phys.} {B 70} {2004} {094514}.
\bibitem{FeSc1}
\Name {Y. Kamihara, T. Watanabe, M. Hirano, H. Hosono}
\REVIEW {J. Am. Chem. Soc.} {130} {2008} {3296}.
\bibitem{FeSc2}
\Name {X. H. Chen, T. Wu, G. Wu, R. H. Liu, H. Chen, D. F. Fang}
\REVIEW { Nature} {453} {2008} {761}.
\bibitem{FeSc3}
\Name {G. F. Chen, Z. Li, D. Wu, G. Li, W. Z. Hu, J. Dong, P. Zheng, J. L. Luo, \and N. L. Wang}
\REVIEW {Phys. Rev. Lett.} {100} {2008} {247002}.
\bibitem{FeSc4}
\Name {Z.-A. Ren, G.-C. Che, X.-L. Dong, J. Yang, W. Lu et al}
\REVIEW {Europhys. Lett.} {83} {2008} {17002}.
\bibitem{FeSc5}
\Name {M. Rotter, M. Tegel, D. Johrendt}
\REVIEW {Phys. Rev.Lett} {101} {2008} {107006}.
\bibitem{FeSc6}
\Name {K. Sasmal, B. Lv, B. Lorenz, A. M. Guloy, F. Chen, Y.-Y. Xue, \and C.-W. Chu}
\REVIEW {Phys. Rev. Lett.} {101} {2008} {107007}.
\bibitem{FeSc7}
\Name {Ni. N, A. Thaler, J. Q. Yan, A. Kracher, E. Colombier, S. L. Bud'ko, P. C. Canfield}
\REVIEW{Phys. Rev.} {B 82} {2010} {024519}.
\bibitem{FeSc8}
\Name {Y. Mizuguchi, F. Tomioka, S. Tsuda, T. Yamaguchi, \and Y. Takano}
\REVIEW {Appl. Phys. Lett.} {93} {2008} {152505}.
\bibitem{FeSc9}
\Name {M. H. Fang, H. M. Pham, B. Qian, T. J. Liu, E. K. Vehstedt, Y. Liu, L. Spinu, \and Z. Q. Mao}
\REVIEW {Phys. Rev.} {B 78} {2008} {224503}.
\bibitem{FeScTh1}
\Name {L. de'Medici, G. Giovannetti, \and M. Capone}
\REVIEW {Phys. Rev. Lett. } {112} {2014} {177001}.
\bibitem{FeScTh2}
\Name {M. J. Calder\'{o}n, L. de'Medici, B. Valenzuela, \and E. Bascones}
\REVIEW {Phys. Rev.} {B 90} {2014} {115128}.

\end{thebibliography}
\end{document}